\documentclass{aa}

\usepackage{graphicx}

\usepackage{txfonts}

\usepackage{url}
\usepackage{hyperref}
\hypersetup{
    final=true,
    pageanchor=true,
    colorlinks=true,
    breaklinks=true,
    linkcolor=blue,
    citecolor=blue,
    urlcolor=blue, 
    pdfpagemode=UseNone,
    pdftitle={aa42199-21},
    pdfauthor={H. Balthasar},
    pdfsubject={The Sun and the heliosphere},
    pdfkeywords={Sun: sunspots -- Sun: magnetic fields -- Sun: spectroscopy}}
\usepackage{amssymb}
\usepackage[usenames, dvipsnames]{color}
\usepackage{graphicx}
\usepackage{natbib}
\usepackage[rightcaption]{sidecap}
\usepackage{txfonts}
\usepackage{units}
\usepackage{url}
\usepackage{pifont}
\usepackage[modulo, switch]{lineno}
\usepackage{ulem}
\usepackage{subcaption}
\usepackage{caption}

\begin{document}

   \title{An unusual velocity field in a sunspot penumbra}

   \author{
          H.\ Balthasar\inst{1},
          C.\ Denker\inst{1},
          A. Diercke\inst{3,1,2},
          S.\ J.\ Gonz\'alez\ Manrique\inst{4,5},
          C.\ Kuckein\inst{4,5},
          J.\ L\"ohner\ B\"ottcher\inst{6,3}, \\
          R.\ E.\ Louis \inst{7,1},
          M.\ Sobotka \inst{8},
          \and
          M. Verma\inst{1}
          }

   \institute{Leibniz-Institut f\"ur Astrophysik Potsdam (AIP), 
              An der Sternwarte 16, D-14482 Potsdam, Germany\\
              \email{hbalthasar@aip.de}
         \and
             Institut f\"ur Physik und Astronomie, Universit\"at Potsdam, Germany 
         \and 
             Institut f\"ur Sonnenphysik (KIS), George-K\"ohler-Allee 401a, D-79110 Freiburg, Germany
         \and    
            Instituto de Astrof{\'\i}sica de Canarias, V{\'\i}a L{\'a}ctea s/n, 38205 La Laguna, Tenerife, Spain
         \and   
            Departamento de Astrof{\'\i}sica, Universidad de La Laguna, 38205 La Laguna, Tenerife, Spain
         \and   
            Hansa-Gymnasium, F{\"a}hrwall 19, D-18439 Stralsund, Germany
         \and
             Udaipur Solar Observatory, Physical Research Laboratory, Udaipur, India 
         \and
             Astronomical Institute of the Czech Academy of Sciences, Fri{\v c}ova 298, 25165 Ond{\v r}ejov, 
             Czech Republic 
             }

   \date{Received 10/09/2021; 30/10/2025}

        \authorrunning{Balthasar et al.}
  \titlerunning{Unusual penumbral velocities}

  \abstract
   {The photospheric Evershed flow is normally oriented radially outward, yet sometimes opposite velocities 
   are observed not only in the chromosphere but also in the photospheric layers of the penumbra.}
   {We study the velocity field in a special case of an active region with two mature sunspots, 
   where one of them formed several days later than the main one.
   Between the two spots, flux emergence is still ongoing influencing the velocity pattern.}
   {We observed the active region NOAA 12146 on August 24, 2014, with the GREGOR Fabry-P\'erot Interferometer
   (GFPI) and the Blue Imaging Channel (BIC) of the GREGOR solar telescope at Observatorio del Teide 
   on Tenerife. Context data from the Helioseismic and Magnetic Imager (HMI) onboard the 
   Solar Dynamics Observatory (SDO) complement the high-resolution data.}
   {In the penumbra of a newly formed spot, we observe opposite Doppler velocity streams of up to 
   $\pm$2\,km\,s$^{-1}${} very 
   close to each other. 
   These velocities extend beyond the outer penumbral boundary and cross also the polarity-inversion line.
   The properties of the magnetic field do not change significantly between these two streams. 
   Although the magnetic field is almost horizontal, we do not detect large transversal velocities 
   in horizontal flow maps obtained with the local correlation technique.}
   {The ongoing emergence of magnetic flux in an active region causes flows of opposite directions 
   intruding the penumbra of a pre-existing sunspot.}


   \maketitle

\section{Introduction}

Regular sunspots exhibit the Evershed flow (EF) in their penumbra. It is an outward flow with a velocity of 
about 2\,km\,s$^{-1}$ \citep[see][]{solanki2003}. The line profiles in the penumbra are asymmetric, 
providing a hint that 
the EF is height-dependent. \citet{SW1980} studied different spectral lines and found decreasing 
velocities with height in the solar atmosphere. \citet{BSW1997} investigated three spectral lines formed 
in different atmospheric layers. The lines were recorded simultaneously, and shifts due to the 
Earth's motion were corrected. 
In deep layers, they found velocities of up to 3\,km\,s$^{-1}$, in medium heights about 1\,km\,s$^{-1}${} and 
in high photospheric layers, the velocities are close to zero on an absolute wavelength scale. 
\citet{Wiehr1995} investigated the asymmetries and explained them as a superposition
of two different velocities. The faster component can be more than 5\,km\,s$^{-1}$.
In chromospheric and higher layers, an inverse EF is observed \citep{StJohn1913}. In contrast to 
the photosphere, 
flow velocities increase with height in the chromosphere. \citet{Alissandrakis} reported peak velocities of 
$-$1.7\,km\,s$^{-1}${} in the photosphere, +1.7\,km\,s$^{-1}${} in the low chromosphere, 
2.5\,km\,s$^{-1}${} in the higher chromosphere, and 15\,km\,s$^{-1}${} in the transition region.
In a recent investigation, \citet{Choudhary_Beck2018} found partly supersonic velocities from the
chromospheric spectral line 
Ca\,{\sc ii}\,854.2\,nm, in contrast to \citet{Tsiropoula2000}, who found subsonic velocities from H$\alpha$.
The inverse EF is explained as a siphon flow by \citet{Choudhary_Beck2018}, 
but for the photospheric EF, an
overturning convection causes the magnetized outflow \citep[see][]{Rempel2015}.

Anomalous flows in the photosphere of sunspot penumbrae were observed in several cases. 
\citet{louis2011} observed three sunspots with locations where Doppler shifts opposite to the 
normal EF were detected. 
Close to the umbra-penumbra boundary, they are interpreted as supersonic downflows.
\citet{KSD2013} investigated a sunspot with bright filaments intruding unusually far into the umbra. 
A feature on the limb-side shows blueshifts in the penumbra and redshifts inside the umbra, 
and the opposite behaviour was found on the center side.
\citet{Louis2014} observed a blueshifted feature
embedded in the limbside penumbra with a velocity of 1.6\,km\,s$^{-1}$. The magnetic field in 
this feature was almost horizontal, and the feature occurred in the prolongation of a light bridge. 
It was accompanied by a redshifted 
area next to it. \citet{Louis2020} investigate an atypical light bridge producing a blueshift which 
reaches into the limb-side penumbra, where the normal EF exhibits a redshift. Another case of 
blueshifts in the limb-side 
penumbra, related to an umbral filament, was investigated by \citet{Guglielmino2019}.
\citet{B_etal2016}
found redshifts in a center-side penumbra when there was an arch filament system (AFS) between the spot and 
the neighboring pores. Some footpoints of the AFS were located in the penumbra. AFS occur when new flux is 
emerging, and matter is streaming downward along the magnetic field lines in the arch filaments.
\citet{Siu-Tapia2017} reported a case where in a part of the penumbra the flow was inward, i.-e. reversed 
with respect
to the regular EF, and called it counter EF. They used data recorded with Hinode SOT/SP in the spectral lines 
Fe\,{\sc i}\,630.15\,nm and Fe\,{\sc i}\,630.25\,nm which form in mid-photospheric layers. 
The magnetic field 
is more horizontal at the location of the counter EF than in the areas of normal EF. 
With numerical simulations, \citet{Siu-Tapia2018} interpreted the normal EF originating from overturning 
convection, and at a few occasions, a counter EF caused by a siphon flow occurs for limited time intervals.
\citet{Murabito2018} studied twelve active regions with 17 spots (ten preceding, seven following)
during the formation of a penumbra. In eleven cases, they found counter EFs before the penumbrae 
became detectable. A more detailed statistical analysis on how often counter EFs appear 
is presented by \citet{CastellanosDuran2021}. 
They find that
81 out of 97 active regions show counter EFs, and they find a median of six cases of counter EFs 
per active region.
Many cases are related to light bridges or deviations of the vertical magnetic field from the surroundings.
Flows opposite to the normal EF also are observed before a penumbra is formed or restored, 
see \citet{Romano2020} and \citet{Garcia_Rivas2024}.

In this work, we investigate an active region with newly emerging magnetic flux next to a pre-existing sunspot 
and study especially the velocity field related to the new flux. We focus on an unusual velocity 
field in this sunspot group.

   \begin{figure}
   \centering
   \includegraphics[scale=0.45, clip=true]{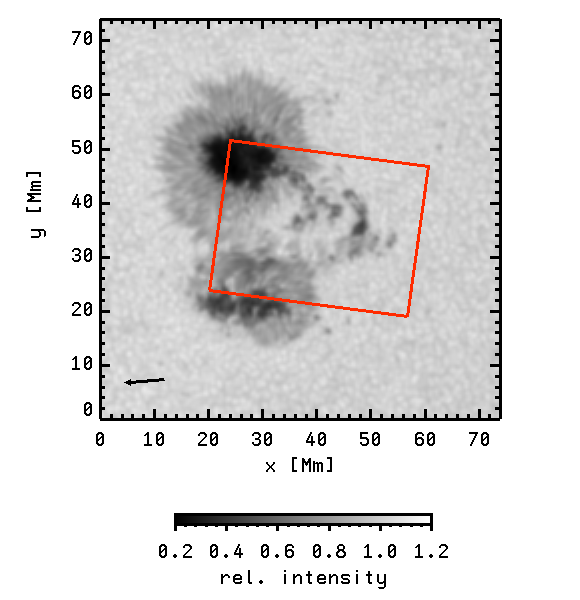}
      \caption{White-light image of the active region obtained with HMI at 09:48 UT on August 24, 2014.
      The red box marks the field of view of GFPI. The black arrow points to the solar disk center.
                        }
         \label{FigHMI_Int}
   \end{figure}

\section{Observations}

We observed active region NOAA 12146 with the GREGOR solar telescope \citep{GREGOR} and the GREGOR
Fabry-P\'erot Interferometer \citep[GFPI;][]{Denker2010, GFPI} 
in spectroscopic mode from 09:43 till 09:59 UT on August 24, 2014.
The region was located at 10$^\circ$ North and 20$^\circ$ West. An overview of the active region is 
displayed in Fig.~\ref{FigHMI_Int}.
We selected the spectral line Fe\,{\sc i}\,709\,nm which does not split in the magnetic field. 
This line has an excitation potential of 4.23\,eV and probes 
medium heights in the photosphere. We recorded 24 positions along the line profile with a single 
exposure time
of 45\,ms and accumulated eight images per wavelength step. A scan through the line profile took 
about 31\,s. 
The step width in wavelength was 2.363\,pm, and the pixel size of the GFPI camera 
corresponds to 0\farcs{}079.
To get the wavelength of 709~nm to the GFPI, we had to replace the standard pentaprism dividing 
the light between GFPI 
and the GREGOR Infrared Spectrograph \citep[GRIS;][]{GRIS2012} by a three-mirror system,\footnote
{In the pentaprism, the deflected beam passes about 22\,cm through glass with a focus extension of 
roughly 8\,cm. Thus the pentaprism cannot simply be replaced by one or two mirrors because the setup 
of the GFPI cannot be moved as a whole. Instead, the light is taken 
out of the main beam before the position of the pentaprism, and two more mirrors are needed to fold 
the beam again 
into the optical axis of the GFPI. The arrangement of the mirrors minimizes polarization effects of 
the mirrors.}
thus we could not use GRIS in parallel.
After correction for dark currents and flatfield, the images were reconstructed with  
Multi Object Multi Frame 
Blind Deconvolution \citep[MOMFBD;][]{MOMFBD2002, MOMFBD2005} using eight images per wavelength step. 
The images from the continuum channel, which were taken strictly simultaneously, were used to align 
the images along 
the spectral scan after determining the instrumental alignment using pinhole and target recordings. 
In the next step, 
the instrumental blue-shift of the GFPI was corrected in the data. Then we removed the transmission 
curve of the prefilter.  
Finally, we corrected the internal image rotation of the GREGOR telescope within the series of 
30 scans through the 
line profile. The derotator was installed only after our observations. These reduction routines are
available in the sTools data processing pipeline \citep{sTools}.

Velocities were determined in two different ways, either by a polynomial fit of fourth order or by the 
Fourier-method described by \citet{FFTSchlich2000}. The results differ in some details. The polynomial fit 
yields the position of the line core (minimum), while the Fourier-method represents the position of the line 
as a whole. Differences between the methods are due to line asymmetries, and because of the line core 
forming in higher layers of the photosphere than the rest of the line. Nevertheless, the global view 
is quite similar.
The zero-point reference is the mean value in a vertical stripe
on the right side of the GFPI-part in Fig.~\ref{FigHMI_Int} which represents the average of granular 
and intergranular velocities. We note that this reference is not necessarily an absolute one.

\begin{figure*}
   \centering
\includegraphics[scale=0.50, clip=true]{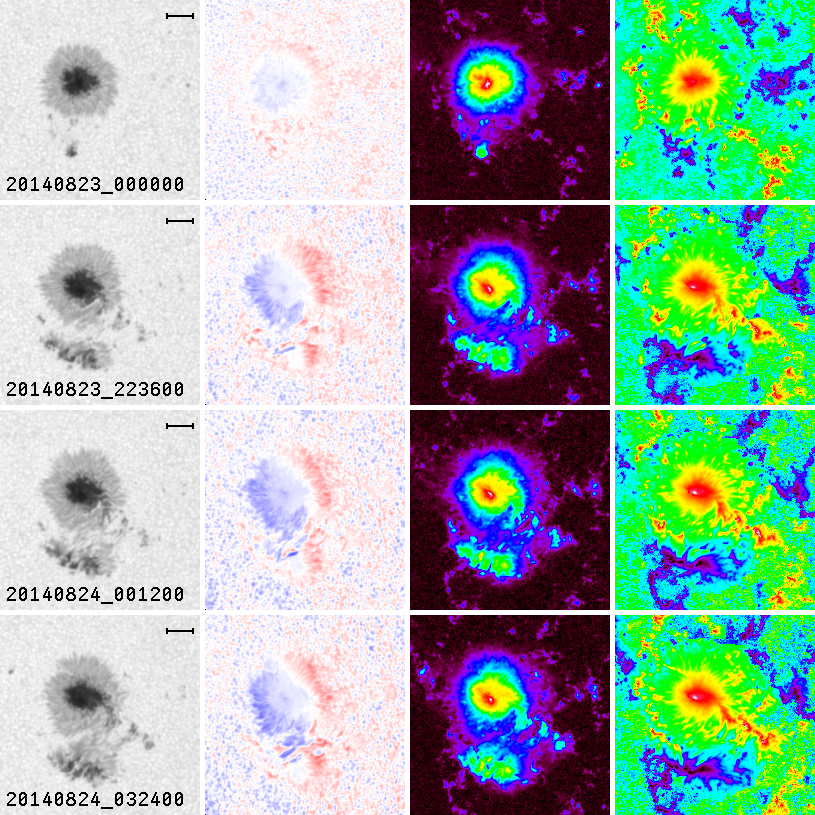}
\caption{HMI-data at four time steps. Each of these time steps is displayed in a row. 
Left column: intensity, column 2: HMI-velocities,
column 3: total magnetic field strength, and right column: magnetic inclination.
Velocities are scaled in the range $\pm$2.5\,km\,s$^{-1}$, the magnetic field strength is clipped at 3000\,G,
and the range of inclination is 0 -- 180$^\circ$. The bar in the upper right corner of the intensity images 
indicates a length of 10\,Mm. An animation of this figure from 00:00\,UT on August 23 till 23:48\,UT on August 24 
is available online.
}
              
         \label{Fig_HMI_multi}
   \end{figure*}

Context images were obtained at GREGOR with two pco4000 CCD-cameras mounted at the Blue Imaging Channel 
\citep[BIC;][]{GFPI}, first mentioned in \citet{Denker2010},
using a G-band (430\,nm) and a blue continuum (450\,nm) filter. We took repetitive bursts of 100 images with 
an exposure time
of 5\,ms in G-band and 4\,ms in the blue continuum. The series of bursts were started when also the 
GFPI-recording 
was started and stopped when the GFPI recording finished. We obtained 32 bursts in G-band and 28 in 
blue continuum. 
The two cameras were controlled by two different computers.
The images were corrected for dark current and flatfield, 
and each burst was processed with the Kiepenheuer-Institute Speckle Interferometry Package 
\citep[KISIP;][]{KISIP1, KISIP2}.
Thus, each burst results in a reconstructed image.
An example for a blue continuum image is shown in  
Fig.~\ref{FigBig}\,a. We select the blue continuum
images because they exhibit a higher granular contrast compared to G-band images.

To get information about the magnetic field, we downloaded data obtained with the Helioseismic and Magnetic 
Imager \citep[HMI;][]{hmi} onboard of the Solar Dynamics Observatory \citep[SDO;][]{Pesnell2012}.
We use the magnetic vector field, the Doppler velocity and the continuum intensity. 
The magnetic field vector was derived by the HMI-team with their vector magnetic field pipeline
\citep{Hoeksema2014} applying the Very Fast Inversion of the Stokes Vector 
\citep[VFISV;][]{VFISV, Centeno2014}. The pipeline includes the `Minimum Energy method' to solve the magnetic 
180$^\circ$ ambiguity. We selected the data taken at 09:48~UT which are the only ones 
in the 720\,s series obtained during the observation period at GFPI.
For the alignment, we compare the continuum images from GFPI and HMI, and cut out the common range 
from the HMI-image (red box in Fig.\,\ref{FigHMI_Int}). Then we interpolated the HMI-image to the same pixel 
size as the GFPI-image. Intensity contour lines of both images agree on average better than a distance 
of 0\farcs{}3 on the Sun. Note that the original HMI-pixel size is 0\farcs{}5.

   \begin{figure}
   \centering
      \begin{subfigure}[b]{10cm}
      \includegraphics[scale=0.45, clip=true]{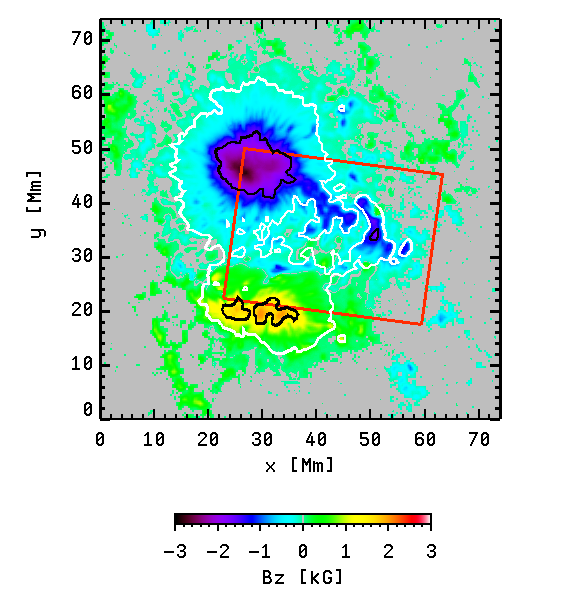}
      \label{fig: a}
      \end{subfigure}
      \begin{subfigure}[b]{10cm}
      \includegraphics[scale=0.45, clip=true]{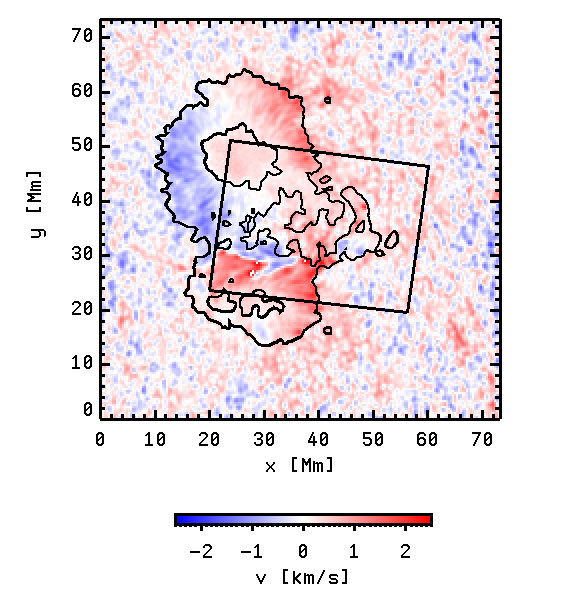}
      \label{fig: b}
      \end{subfigure}     
\caption{Upper: Vertical component of the magnetic field recorded by HMI
at 09:48\,UT on August 24. 
Black and white contours indicate the boundaries of umbra 
      and penumbra according to Fig.~\ref{FigHMI_Int}. Values are suppressed where the total magnetic field 
      strength is below 150\,G. 
      Lower: Corresponding Doppler-velocity from HMI.
         }              
      \label{Fig_Bz}
   \end{figure}

  \section{Results}
  
This active region developed very quickly. On August 21, it consisted of a single regular spot of negative 
polarity. 
In the following, we describe the development of intensity, Doppler velocity, magnetic field strength, 
and inclination taken 
from HMI between 00:00\,UT on August 23 and 23:48\,UT on August 24.
Fig.~\ref{Fig_HMI_multi} shows an example of the HMI-images, and an animation is available online.
At 00:00 UT on August 23, a few hours after the meridian passage, a single pore of positive 
polarity was visible south of the main spot.
Then more positive flux in form of pores emerged. Negative flux separated from the main spot
and moved toward the positive region, where it was cancelled, perhaps by submerging flux loops. 
Finally, the positive pores merged and formed the second sunspot. On August 24, the two 
penumbrae came into contact, but separated later again. High Doppler-velocities appeared for certain periods
in the area between the major spots and disappeared again. 
East of the two major spots was another small spot with an irregular penumbra, and several pores appeared
between the main and the small spot. Most of these pores moved towards the main spot and merged with it.
Next to the two major spots, the emergence of magnetic flux was still ongoing on August 25. 
The configuration of the vertical magnetic field is shown in Fig.~\ref{Fig_Bz}.
From this vertical component of the magnetic field, we derive the polarity inversion line (PIL). 
Between August 22 and 29, the group caused 17 C-class flares, and on August 25, 
two M-class flares occurred in this group. 
On August 24, three C-class flares happened at 00:08, 07:26 and 11:50\,UT.

\begin{figure*}[htp]
\centering
\begin{subfigure}[b]{0.40\textwidth}
\includegraphics[scale=0.25, trim=1.0cm 1.0cm 1.0cm 0.0cm, clip=true]{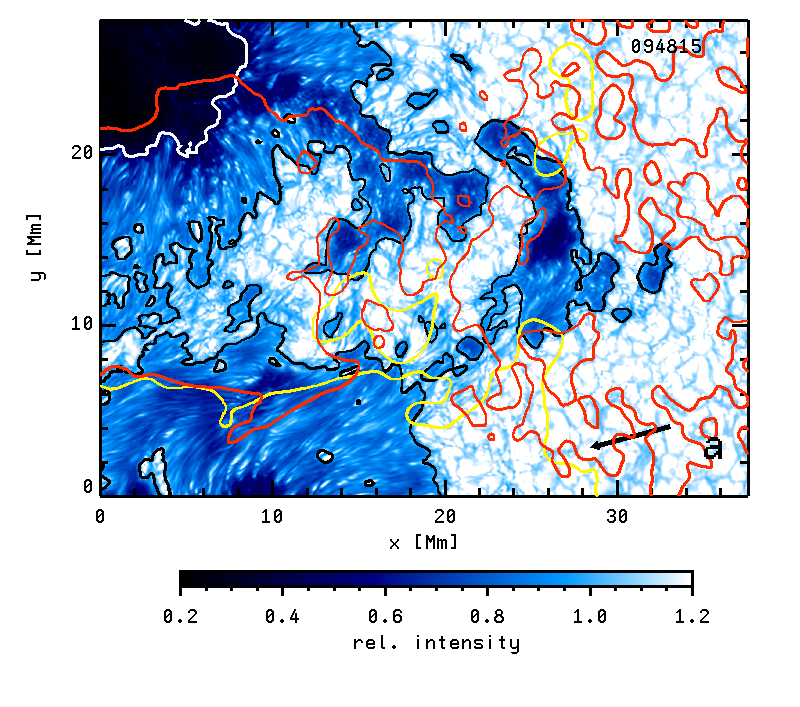}
\label{fig: a}
\end{subfigure}
\begin{subfigure}[b]{0.40\textwidth}
\includegraphics[scale=0.25, trim=1.0cm 1.0cm 1.0cm 0.0cm, clip=true]{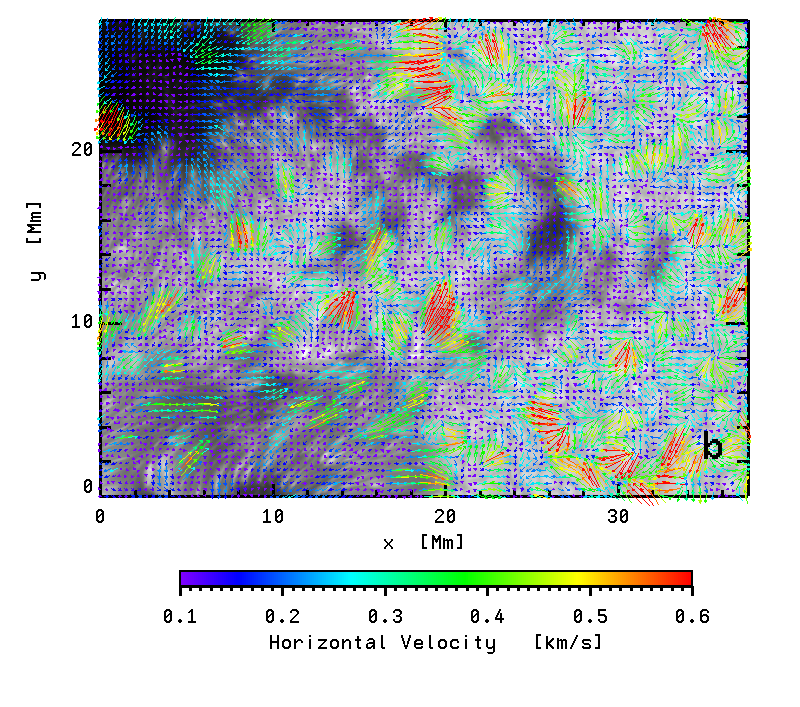}
\label{fig: b}
\end{subfigure}
\begin{subfigure}[b]{0.40\textwidth}
\includegraphics[scale=0.25, trim=1.0cm 1.0cm 1.0cm 0.0cm, clip=true]{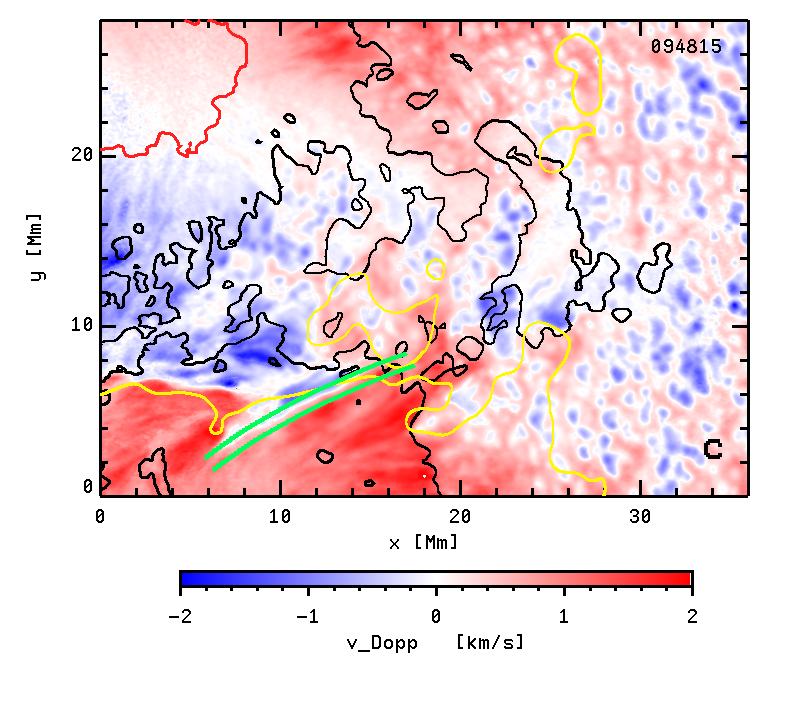}
\label{fig: c}
\end{subfigure}
\begin{subfigure}[b]{0.40\textwidth}
\includegraphics[scale=0.25, trim=1.0cm 1.0cm 1.0cm 0.0cm, clip=true]{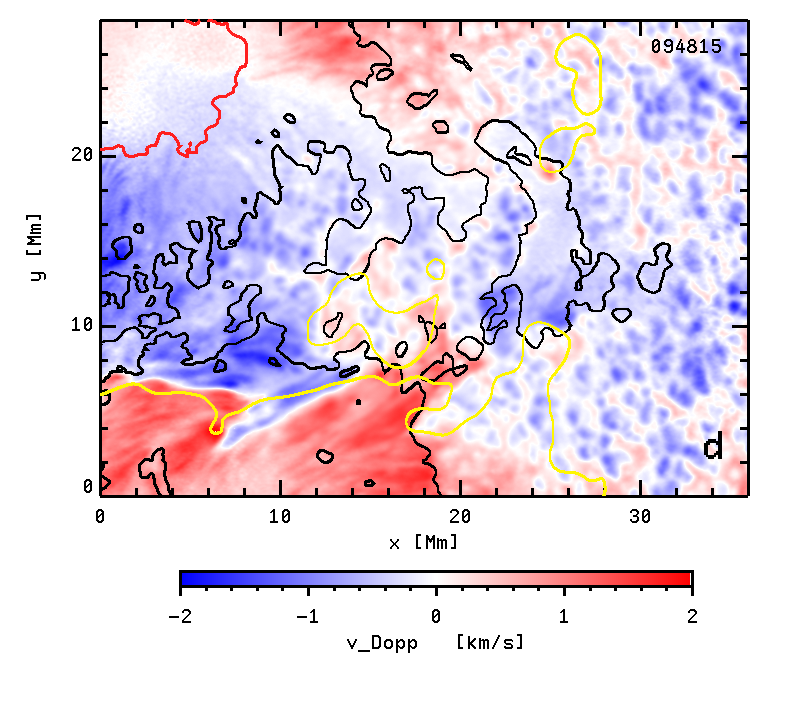}
\label{fig: d}
\end{subfigure}
\begin{subfigure}[b]{0.40\textwidth}
\includegraphics[scale=0.25, trim=1.0cm 1.0cm 1.0cm 0.0cm, clip=true]{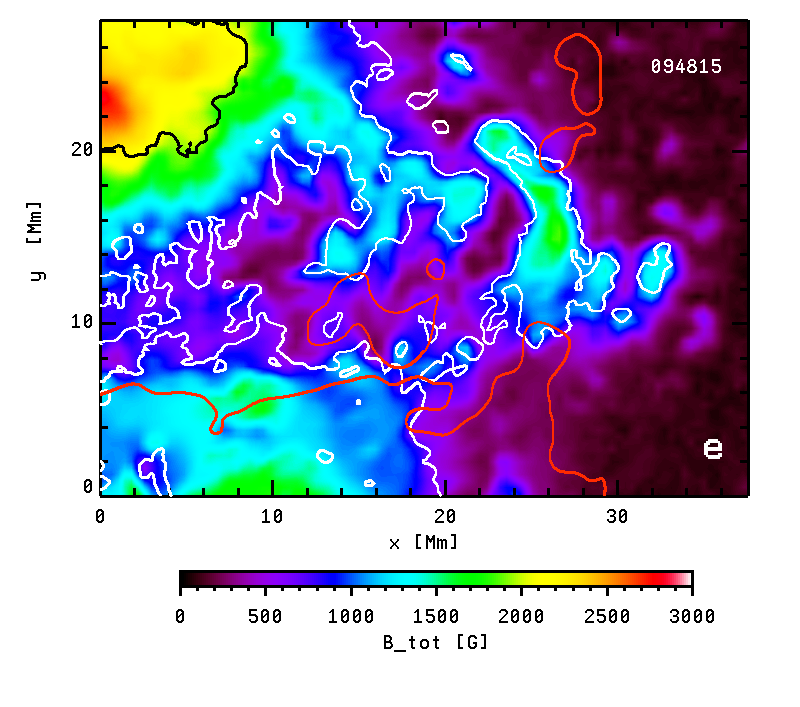}
\label{fig: e}
\end{subfigure}
\begin{subfigure}[b]{0.40\textwidth}
\includegraphics[scale=0.25, trim=1.0cm 1.0cm 1.0cm 0.0cm, clip=true]{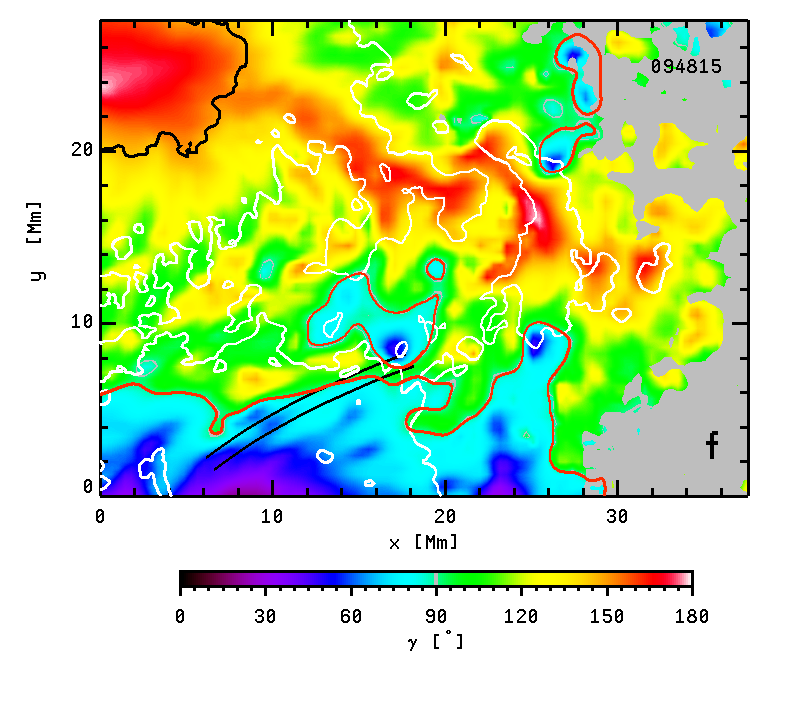}
\label{fig: f}
\end{subfigure}
\caption{a: Speckle restored blue continuum image of the investigated region obtained at GREGOR,
      restricted to the area recorded with the GFPI.    
      The red contours mark the zero-velocity lines and the yellow line the PIL.
      Black contours outline the outer penumbral boundary and white the umbral-penumbral boundary.
      The arrow points to disk center (note image rotation compared to Fig.\,\ref{FigHMI_Int}). 
      b: horizontal velocity map obtained by LCT. Color and length of the arrows indicate the velocity 
      in a spectral scale from black to red according to the color bar. The average blue continuum image 
      serves as background.
      c and d: Velocity maps, derived from polynomial fits (c) and from the Fourier-method (d) 
      applied to GFPI-data.
             Red and black contours mark the 
             umbra and penumbra according to the GFPI intensity map. The yellow line indicates the PIL.
             Here, trajectories of high velocities shown in Fig.~\ref{Fig_graph} 
             are marked in light blue.
      e:  Total magnetic field strength.       
      f:      Inclination of the magnetic field in the local frame of reference, 
              taken from HMI. 
              Black and white contours indicate the boundaries 
              of umbra and penumbra in the GFPI-data.
              Values are set to 90$^\circ$ where the total magnetic 
              field strength is below 150\,G. The violet line marks the PIL.
              The trajectories are marked in panel f in black.
              An animation of panels c and d is available online.      
       }
\label{FigBig}
\end{figure*}

The most prominent result of our observations is the velocity field in the penumbra of the newly formed sunspot.
The 30 scans through the line profile give very similar results, that means that the feature was persistent for 
at least 16 minutes. Thus, we display only maps from the scan that 
started at 09:48:15\,UT, when we have nearly simultaneous data from HMI.
Velocities of up to 2\,km\,s$^{-1}${} in opposite directions appeared very close to each other, as shown in  
Figs.~\ref{FigBig}\,c and \ref{FigBig}\,d. Some blueshifts are more prominent in the Fourier-method 
(Fig.~\ref{FigBig}\,d), probably because the Fourier-method is more sensitive to the whole line profile 
stemming from deeper layers than the line core which is relevant for the polynomial fits.
The flows in both directions cross the PIL, i.-e. they are not 
separated by the PIL. At this position on the disk, we expected a velocity away from us according to the 
EF for the newly emerging sunspot, while very low velocities were expected for the nearby penumbra of 
the main spot. In fact, we observe blueshifts in this area, and the space between the two spots is 
also dominated by 
blueshifts. These features are also seen in the HMI Doppler velocities, but at lower resolution. 

Normally, penumbral filaments are oriented radially, but as can be seen in Fig.~\ref{FigBig}\,a, they 
deviate significantly from the radial direction, and the outer ends point more towards disk center. Similar 
is the curvature in the penumbra of the newly formed spot, especially at the locations where the opposite 
velocities occur. Here, they point away from disk center. The two major spots do not have a common penumbra 
during the time of the GFPI observations. They are separated by a narrow belt
between the penumbrae where the intensity is higher than in the penumbra, but which is influenced by the
magnetic field (abnormal granulation). 
Granules are elongated, and intergranular 
lanes are straighter than usual, and their orientation 
is aligned with the nearby penumbral filaments.

To determine the horizontal velocities, we used the BIC blue continuum data and applied the 
local correlation technique (LCT)
as described by \citet{VD2011} and \citet{Verma2016a}. The velocities were averaged for 
$\Delta T  =$ 13 min. For the whole area, we obtained comparably small velocities of up to 0.6\,km\,s$^{-1}$.
The results are displayed in 
Fig.~\ref{FigBig}\,b.
The highest velocities occur in local areas outside the spots which appear grey in Fig.~\ref{FigHMI_Int}, 
and at the umbra-penumbra boundary of the main spot.  
The values along the trajectories marked in 
Fig.~\ref{FigBig}\,c are rather small, less than 0.3\,km\,s$^{-1}${}. Only at 
$x$~=~$14^{\prime\prime}$ and $y$~=~$8^{\prime\prime}$, we find velocities of about 0.5\,km\,s$^{-1}$. 
However, LCT needs moving contrast features,
and a laminar flow along magnetic field lines  without intensity variations will not show up in LCT. 
In general, LCT-velocities are very small in the penumbrae, often even smaller than 0.3\,km\,s$^{-1}$.

   \begin{figure*}
   \centering
 \begin{subfigure}[b]{0.40\textwidth}  
\includegraphics[scale=0.26, trim=0.3cm 0.3cm 0.5cm 0.0cm, clip=true]{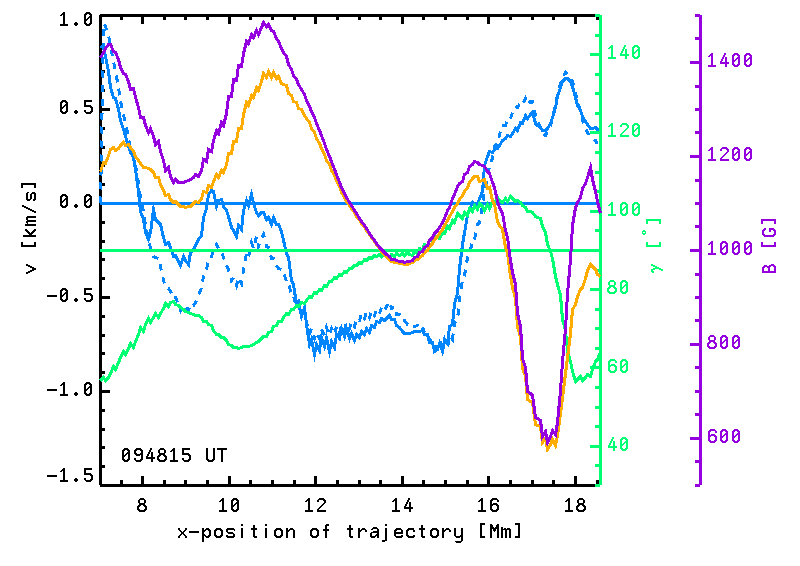}
\label{fig: va}
\end{subfigure}
\begin{subfigure}[b]{0.40\textwidth}
\includegraphics[scale=0.26, trim=0.3cm 0.3cm 0.5cm 0.0cm, clip=true]{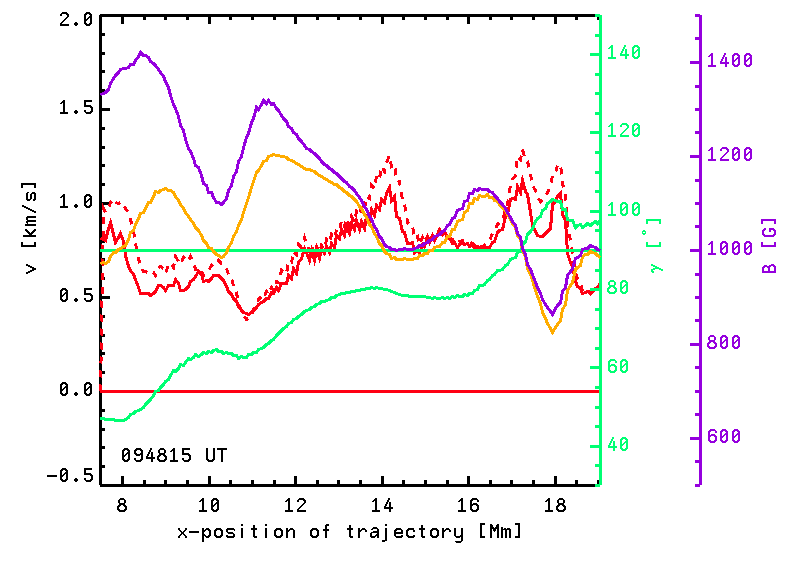}
\label{fig: vb}
\end{subfigure}
       \caption{Left panel: Parameters along the line with maximum blueshifts. 
              Solid blue: Doppler velocities 
              from the polynomial fit, 
              dashed blue: Doppler velocities from the Fourier-method, green: magnetic inclination, 
              violet: total magnetic field strength, and orange: horizontal magnetic field strength.
              Horizontal lines mark the zero reference for velocities (blue) and 90$^\circ$ for the 
              magnetic inclination (green). 
              Right panel: Parameters for the maximum redshift. Here, the velocities are displayed in red. 
              An animation of this figure is available online.
              }
         \label{Fig_graph}
  \end{figure*}

The total magnetic field strength is shown in Fig.~\ref{FigBig}\,e. In the penumbra of the 
newly formed spot it varies between 1000\,G and 2000\,G, but we do not see special structures related
to the trajectories of the high Doppler velocities.
The structure of the vertical magnetic field is not related to the opposite velocities in the penumbra. 
This can be seen in the magnetic inclination map in 
Fig.~\ref{FigBig}\,f.
To investigate the high velocities and their relation to the magnetic field, we extract velocities, 
total magnetic field strength, its horizontal component, and the magnetic inclination along two 
trajectories where the velocities are highest:
the `blue' trajectory, which is the upper one in  
Fig.~\ref{FigBig}\,c, and the `red' trajectory, the lower one in the figure.
The two trajectories have the same length, and the second one is shifted against the first one. 
Because of this, blueshifts do not occur all along the `blue' trajectory, the velocities turn to 
redshifts at both ends of this trajectory. 
We average the quantities at each pixel along the trajectory with that of the pixel above and the one below 
in order not to depend too much on possible inaccuracies at a single pixel. The trajectories are  marked in 
Fig.~\ref{FigBig}\,c. The curves are displayed in Fig.~\ref{Fig_graph}, 
where we select the velocity data obtained at 09:48:15~UT, which are closest in time to the magnetic field 
data from HMI.
At the start of the trajectory shown in the left panel of Fig.~\ref{Fig_graph}, we still have redshifts. 
At $x$-position 8\,Mm, they 
reduce to almost zero, and at $x$ = 11\,Mm the velocities turn into the negative (blueshifted) range.
The velocities do not change much where the inclination crosses the 90$^\circ$ line of the magnetic
inclination. Only at 
$x$ = 16\,Mm, blueshifts turn into redshifts. Here, the magnetic inclination reaches its maximum 
of about 100$^\circ$, 
and the field strength starts to decrease rapidly from more than 1000\,G to about 600\,G. 
In the right panel of Fig.~\ref{Fig_graph} we display the values for the parallel trajectory 
with strong redshifts.
Variations are smoother 
than in left panel of Fig.~\ref{Fig_graph} and all along the trajectory, we deal with redshifts. 
A special 
reaction of the velocities where the magnetic inclination crosses the 90$^\circ$-line is not visible. 
We have a local maximum at this point, but similar local maxima occur on both sides of this point 
along the trajectory. 
The variations of the magnetic field probably indicate that the trajectories are not exactly parallel to the 
magnetic field lines.

Investigating carefully the temporal evolution along the trajectories, some locations show velocity variations. 
Our time series is too short for a period analysis. 
An animation of the velocity variations along the trajectories is provided online.
\citet{Rimmele94} determined a quasi-periodic time scale of 
15 min for variations of the EF, corresponding to the total length of our series.
However, it looks that the variations in our series correspond to standing waves.

\section{Discussion}

The redshifts in the penumbra of the newly formed spot correspond to a regular EF,
but the blueshifts do not. Counter EFs often occur in relation to emerging magnetic flux.
The antiparallel flows investigated in this paper lasted from 08:48\,UT until 
10:36\,UT, somewhat less than two hours. In the movie attached to Fig.\,\ref{Fig_HMI_multi}
one can detect several similar cases of high velocities at narrow locations before and 
after the observation with GFPI. This shows that we deal with highly dynamic processes 
because of the ongoing flux emergence. However, we do not recognize major changes in the 
photosphere related to the mentioned flares.

\citet{B_etal2016}
reported redshifts in a center-side penumbra, which 
were related to the emerging flux of an AFS. The footpoints of the AFS harbor downflows, 
and some of them were located in the penumbra. 
Downflows in the footpoints were also observed by \citet{SGMetal2018}.
In contrast, we observe strong blueshifts intruding into a 
penumbra that should exhibit a redshifted EF. 
Flux emergence was still continuing during our observations in an
area between both spots. An explanation for this velocity field can be that in flux tubes belonging 
to the spot, the flow follows the direction of the regular EF, while the flow is in opposite direction 
in flux tubes 
belonging to the newly emerging flux. Magnetic field lines are more or less parallel, however, the direction of 
the flow is independent from the polarity of the magnetic field. As can be seen in  
Figs.~\ref{FigBig}\,b,  and \ref{Fig_graph}, 
the magnetic inclination at the location of the high velocities is close to 90$^\circ$, i.-e. the magnetic 
field is almost horizontal, and the opposite velocity directions do not show up as a difference in 
magnetic inclination. The stream follows the magnetic field lines as shown by \citet{BBCS2003}, thus
the true velocities can be up to 5\,km\,s$^{-1}$, but still less than the speed of sound 
($\approx7$\,km\,s$^{-1}$).

Active region NOAA 12146 is included in the investigation of \citet{CastellanosDuran2021} for which they
detected four cases of counter EFs,
but not on August 24. 
Counter EFs can also appear in orphan penumbra, as reported by \citet{CastellanosDuran2025}.
However, the penumbra in our case is not an orphan one.
In our observations we cannot see any light bridge related to the blueshift, 
which extends even into 
the quiet Sun, nor do we see a local deviation of the vertical magnetic field from the surrounding areas. 
On the other hand, our case also differs from that of \citet{Siu-Tapia2017}, which is not related to 
newly emerging magnetic flux.

A numerical simulation is presented by \citet{Chen_etal2017} where they found counter EF in an
already formed penumbra and ongoing flux emergence next to it. The counter EF covers almost the whole penumbra 
on one side of the spot, and the inflow becomes supersonic. 
\citet{Siu-Tapia2018} investigated another MHD simulation with a narrower grid. Here, counter EFs appear at 
limited locations, and they last only for a few hours. The inflow is also supersonic close to the umbra,
but with a peak speed of 12\,km\,s$^{-1}$ instead 15\,km\,s$^{-1}$ in case of \citet{Chen_etal2017}.
In contrast, in our observations we find no hints of supersonic flows.
\citet{Siu-Tapia2018} mentioned that the used narrow grid is required for studies of the EF.

The proximity of pre-existing and newly emerging flux forces deviations from the radial direction of
penumbral filaments. Nevertheless, the magnetic field lines are more or less parallel for both 
kinds of flux tubes as we do not detect significant differences of the magnetic properties in the areas of 
different velocities. The magnetic field configuration has to re-arrange during ongoing emergence 
of flux, and that might cause fast flows in both directions.
From our data, we cannot distinguish if the flows, especially the blueshifted ones, are caused by 
overturning convection 
or by a siphon mechanism. According to \citet{Rempel2015}, the normal EF is due to the first mechanism, 
but for the blueshifted flow, we can only speculate.

\section{Conclusions}

Our observations show that in complex sunspot groups with ongoing magnetic flux emergence, 
a flow of matter follows magnetic field lines,
but the direction of the flow is not correlated with the magnetic polarity. Magnetic flux tubes can be 
almost parallel and very close to each other, but harbor flows of opposite direction. 
The direction of the flows depend on where the flux tubes are rooted, in the pre-existing part of the active 
region or in the newly emerging flux. 
We observe the counter EFs inside the new spot's penumbra after its formation, thus this case 
differs from that of \citet{Romano2020} who observed the counter EF between disappearance and reformation 
of the penumbra in the corresponding region, or that of \citet{Schliche2011} and 
\citet{Murabito2016, Murabito2018}, 
where the counter EF was observed before the formation of the penumbra, 
and after the formation, the motion changed to the regular EF. Obviously, in our case, 
the high EFs and counter EFs are related to the emergence of new magnetic flux.

The difference between the two methods to determine velocities indicates 
that flows are stronger in deep photospheric layers than in higher ones. Velocities of the flow can be quite 
high, but we have no indication that they reach the speed of sound.
We assume that these flows are laminar, at least they are not related with temporal intensity fluctuations, 
thus they do not show up in the LCT-analysis.

Complex sunspot groups, especially those which exhibit counter EFs, deserve more attention in the
future. New instruments such as the instrument suites of DKIST \citep{DKIST2016, Rimmele2020} 
and Solar Orbiter \citep{SolarOrbiter}, 
and, in future, of the EST \citep{EST}  
will enable observers to engage in such investigations.

\section{Data availability}

Animations attached to Figs. \ref{Fig_HMI_multi}, \ref{FigBig}  and \ref{Fig_graph} 
are available at

https://www.aanda.org

\noindent The GFPI-data are available upon request from AIP, contact the instrument PI, C.~Denker 
(cdenker@aip.de).

\begin{acknowledgements}
   The 1.5-meter GREGOR solar telescope was built by a German 
consortium under the leadership of the Institute for Solar Physics 
in Freiburg (former Kiepenheuer-Institute) with the Leibniz Institute for Astrophysics Potsdam, 
and the Max-Planck-Institute for 
Solar System Research in G\"ottingen as partners, and with contributions by the Instituto de 
Astrof{\'\i}sica de Canarias and the Astronomical 
Institute of the Academy of Sciences of the Czech Republic. 
SDO/HMI data are provided by the Joint Science
Operations Center -- Science Data Processing. 
This work was partly supported by a joint DFG\,--\,GACR science grant under 
DE~787/5-1--18-08097J. 
CK acknowledges grant RYC2022-037660-I and SJGM
grant RYC2022-037565-I, both funded by MCIN/AEI/10.13039/501100011033 and by ''ESF Investing in your future.
Financial support from grant PID2024-156538NB-I00, funded by MCIN/AEI/ 10.13039/501100011033'' 
and by ''ERDF A way of making Europe'' is gratefully acknowledged by SJGM.
 \end{acknowledgements}

\bibliographystyle{aa}
\bibliography{aa42199-21}

\end{document}